\documentstyle[prl,aps,preprint,tighten,epsfig]{revtex}
 \newcommand {\bi} {\bibitem}
 \newcommand {\be} {\begin{equation}}
\newcommand {\bea} {\begin{eqnarray} \nonumber }
\newcommand {\ee} {\end{equation}}
\newcommand {\eea} {\end{eqnarray}}

\newcommand {\la} {\langle}
\newcommand {\ra} {\rangle}
 \newcommand {\al} {\alpha}
\def\(({\left(}
\def\)){\right)}
\def\[[{\left[}
\def\]]{\right]}
\newcommand{\nn}{\nonumber}

\def \intk{\int {d^dk \over (2 \pi)^d}}

\begin{document}
\draft


\title{Thermodynamics of glasses: a first principle computation}
\author{Marc M\'ezard\cite{Marc}}
\address{Laboratoire de Physique Th\'eorique de l'Ecole
Normale Sup\'{e}rieure \footnote{Unit\'e propre du CNRS,  associ\'ee
 \`a\ l'Ecole
 Normale Sup\'erieure et \`a\ l'Universit\'e de Paris Sud}\\
24 rue
 Lhomond, F-75231 Paris Cedex 05, (France)}
\author{Giorgio Parisi\cite{Giorgio}}
\address{Dipartimento di Fisica and Sezione INFN,\\
Universit\`a di Roma ``La Sapienza'',
Piazzale Aldo Moro 2,
I-00185 Rome (Italy)}

\date{\today}

\maketitle

\begin{abstract}
We propose a first principle computation of the thermodynamics of simple
fragile glasses starting
from the two body interatomic potential. A replica formulation translates this problem into
that of a gas of interacting molecules, each molecule being built of $m$ atoms, and having a 
gyration radius
(related  to the cage size) which vanishes at zero temperature. We use a small cage expansion, valid
at low temperatures, which allows to compute the cage size, the specific heat (which
follows the Dulong and Petit law), and  the configurational entropy.
\end{abstract}

Take a three dimensional classical system consisting of $N$  particles, interacting by
pairs through a short range potential. Very often this system will undergo, upon cooling or
upon density increasing, a solidification into an amorphous solid state- the glass state.
The conditions required for observing this glass phase is the avoidance of crystallisation,
which can always  be obtained  through a fast enough quench (the meaning of 'fast' depends very
much of the type of system) \cite{glass_revue}. There also exist potentials
which naturally present some kind of frustration with respect to the crystalline structures
and therefore solidify into glass states, even when cooled
 slowly- such is the case for instance of binary
mixtures of hard spheres, soft spheres, or Lennard-Jones particles with 
appropriately different radii. These have been studied a lot in recent numerical simulations
\cite{gpglass,BK1,BK2,FRAPA,ColPar}.

Our aim is to compute the thermodynamic properties of this glass phase, using the statistical 
mechanical approach, namely starting from the microscopic Hamiltonian.  The general framework of our 
approach  finds its roots in old ideas of Kauzman \cite{kauzman}, Adam and Gibbs 
\cite{AdGibbs}, which received a boost when Kirkpatrick, Thirumalai and Wolynes underlined the 
analogy between structural glasses and some generalized spin glasses \cite{KiThiWo}.  In this 
framework, which should provide a good description of fragile glass-formers, the glass transition, 
measured from dynamical effects, is associated with an underlying thermodynamic transition at the 
Kauzman or Vogel-Fulcher temperature $T_K$.  This ideal glass transition is the one which should be 
observed on infinitely long time scales \cite{glass_revue}.  This transition is of an unusual type, 
since it presents two apparently contradictory features: 1) The order parameter is discontinuous
at the transition: 
defining the order parameter as the inverse radius of the cage seen by each particle, it jumps 
discontinuously from $0$ in the liquid phase to a finite value in the glass phase.  2) The 
transition is continuous (second order) from the thermodynamical point of view: the free energy is 
continuous, and the transition is signalled by a discontinuity of the specific heat which jumps from 
its liquid value above $T_s$ to a value very close to that of a crystal phase below.  These 
properties are indeed observed in generalized spin glasses
\cite{GrossMez}. The problem of the existence or not of a diverging correlation
length
is still an open one \cite{corr_length}.

 This analogy is suggestive, but it also hides some very basic differences, like the
fact that spin glasses have quenched disorder while structural glasses do not. The
recent discovery of some generalized  spin glass systems without quenched disorder \cite{nodis}
has given credit to the idea that this analogy is not fortuitous.
 The problem was to turn this general idea into a consistent computational scheme allowing
for some quantitative predictions. Important steps in this direction were invented
in \cite{remi,pot}, which showed the necessity of using several copies of the same system in order
to define properly the glass phase. In a previous preliminary study, we used some of these
ideas to estimate the glass temperature, arriving from the liquid phase
\cite{MPhnc}. Here we concentrate instead on the properties of the glass phase itself,
and particularly its properties at low temperatures.

The Hamiltonian of our problem is simply given by:
\be
H=\sum_{1 \le i \leq j \le N} v(x_i-x_j)
\ee
where the particles move in a volume $V$ of a d-dimensional space, and $v$ is an
arbitrary short range potential. We shall
 take the thermodynamic limit $N,V \to \infty$ at fixed density 
$\rho=N/V$. For simplicity, we do not consider here the description of mixtures,
which is presumably an easy generalisation.
The main obstacle to a study of the glass phase is the very description of the
amorphous solid state. In principle one should give the average position of each atom
in the solid, which requires an infinite amount of information. Had we known this information,
we could have added to the Hamiltonian an infinitesimal but extensive 
pinning field which attracts each particle
to its equilibrium position, sending $N$ to infinity first, before  taking the limit of zero 
pinning field.
This is the usual way of identifying the phase transition. In order to get around the problem of
the description of the amorphous solid phase, a simple method has been developed
in the spin glass context-  although one does not know the conjugate field,
 the system itself will
know it, and the idea is to consider two copies (sometimes called 'replicas') of the system,
with an infinitesimal extensive attraction. In the spin glass case this is a very nice method
which allows
to identify the transition  temperature from the fact that the two replicas remain close to
each other in the limit of vanishing coupling \cite{Toulouse,carparsour}.

However this method is too naive and needs to be modified  for the case of glasses.
The reason has to do with the degeneracy of glass states.
This property can be studied in detail in generalized spin glass mean field models
\cite{crisomtap,kurparvir}. For structural glasses, this is a conjecture
which we shall make, on the basis of its agreement with 
 the phenomenology of glasses \cite{ColPar}. 
Let us assume that we can introduce a free energy functional 
$F(\rho)$ 
which depends on the density $\rho(x)$ and on the temperature.  We suppose that at sufficiently low 
temperature this functional has many minima (i.e.  the number of minima goes to infinity with the 
number $N$ of particles).  Exactly at zero temperature these minima coincide with the mimima of 
the potential energy as function of the coordinates of the particles.  Let us label them by an 
index 
$\alpha$.  To each of them we can associate a free energy $F_\al$ and a free energy density 
$f_\al= F_\al/N$. The number of free energy minima with 
free energy density  $f$ is supposed to be exponentially large:
\be
{\cal N}(f,T,N) \approx \exp(N\Sigma(f,T)),\label{CON}
\ee
where the function $\Sigma$ is called the complexity or the configurational entropy (it is the 
contribution to the entropy coming from the existence of an exponentially large number of locally 
stable configurations), which is not defined in the regions $f>f_{max}(T)$ or $f<f_{min}(T)$, where ${\cal 
N}(f,T,N)=0$, and is supposed to go to zero at $f_{min}(T)$, as found in all existing models so far.  In 
the low temperature region the total free energy of the system ($f_{S}$) can be well approximated by 
the sum of the contributions to the free energy of each particular minimum:
\be
Z\equiv e^{-\beta N f_{S}} =\sum_\al e^{-\beta N f_\al} \simeq 
\int_{f_{min}}^{f_{max}} df \ e^{-N[\beta f- \Sigma(f,T)]} \ ,
\label{SUM}
\ee
which shows that the minima which dominate the
sum are those with a free energy density $f^*$
 which minimizes the quantity $\Phi(f)=f-\Sigma(f,T)/\beta$.
The Kauzman temperature $T_K$ is that below which the saddle point sticks
at the minimum: $f^*=f_{min}(T)$. It is the only temperature at which there exists a
thermodynamic singularity.
Another characteristic temperature is the so called dynamical
temperature $T_D$: for $T_D>T>T_K$ the
free energy is still given the fluid solution with constant $\rho$ and at the same time the free 
energy is also given by the sum over the non trivial minima \cite{pot,remi}, and $f^*$ lies inside
the interval $f_{min},f_{max}$. The system may stay in one of the many possible minima. 
The entropy of the system is thus the 
sum of the entropy of a typical minimum and of $\Sigma(f^*,T)$, 
which is the contribution to the entropy 
coming from the exponentially large number of microscopical configurations. It is not obvious
why this is equal to the liquid free energy. In this regime, the 
 time to jump from one minimum to an other minimum is quite large: it is an activated process 
which is controlled by the height of the barriers which separate the different minima.  The 
correlation time will become very large below $T_D$ and for this reason $T_D$ is called the 
dynamical transition point. It is also the mode-coupling 
transition temperature \cite{gotze,fraher,BCKM}.
  The real divergence of the correlation time
appears at $T_{K}$.

In order to get quantitative information on the behaviour of the system it is useful to consider the 
thermodynamics of $m$ replicas which are constrained to stay in the same minimum \cite{remi}; this 
can be done introducing an extensive attraction among replicas which eventually goes to zero.  In 
the same notation as before partition function is:
\be
Z_{m} = \int_{f_m}^{f_M} df \ e^{-N [m \beta f- \Sigma(f,T)]}
\ee
which obviously coincides with the previous one for $m=1$
In the limit 
where $m
\to 1$ the corresponding partition function $Z_m$ is dominated by the correct saddle point $f^*$, 
when the temperature is in the range $T_K<T<T_D$. 
For $T<T_K$, the saddle point $f^*$ sticks at $f^*=f_{min}(T)$ and 
the replicated free energy $F_m=-\log(Z)/(\beta m)$ is maximum at
a value of $m=m^{*}$ smaller than one. One can use expression valid in the 
liquid phase (i.e.  high temperature formulae) to evaluate the free energy $F_m$
at $m<m^{*}$.  We shall write down more explicit 
formulas in our case below.  Notice that the 'replicas' which we introduce here play
a slightly different role compared to 
the ones used in disordered systems: there is no quenched disorder here, and no need to average a 
logarithm of the partition function.  `Replicas' are introduced to handle the problem of the absence 
of description of the amorphous state.  We do not know of any other procedure to characterize an 
amorphous solid state in the framework of equilibrium statistical mechanics.  There is no `zero 
replica' limit, but there is, as in disordered systems, an analytic continuation in the number of 
replicas.  We shall see that this continuation looks rather innocuous. An alternative
method is to introduce a real coupling of
 the system to another system which is thermalized\cite{pot}; this has been
used recently in order to study the glass phase \cite{FRAPA,CarFraPar}

Let us turn to a more explicit implementation of these ideas. The original partition function,
for $N$ undistinguishable particles, is:
\be
Z={1 \over N!} \int \prod_{i=1}^N (d^d x_i) \ 
 \exp\((-\beta \sum_{1 \le i < j \le N} v(x_i-x_j) \))
\label{Z1}
\ee
We introduce $m$ replicas of each particle, and compute $Z^m$, in presence of
 an infinitesimal pinning field which is
an attractive potential between them. This attractive potential
$\phi(x_1,...,x_N)$ should not break the
undistinguishability of all $N$ particles with the same replica index. We
have found it convenient to use the attractive potential:
\be
\exp \((- \beta \phi(x_1,...,x_N) \))= {1 \over N!} \sum_{\{ \pi_a \in S_N \} }
\exp\((-\beta
 \sum_i
\sum_{a,b} w(x^a_{\pi_a(i)}-x^b_{\pi_b(i)}) \))
\ee
where $w$ is a small  attractive potential which is short range
 (the range should be less than the
typical interparticle distance in the solid phase), but its precise form is irrelevant.
We then get a 'replicated' partition function:
\be
Z_m={1 \over N!} \int \prod_{i=1}^N \prod_{a=1}^m (d^d x_i^a) \ 
 \exp\((-\beta \sum_{1 \le i < j \le N} \sum_{a=1}^m v(x_i^a-x_j^a) 
- \beta \sum_{i=1}^N \sum_{a,b=1}^m w(x_i^a-x_i^b)\))
\label{Zm}
\ee

 A finite $w$ gives rise to the formation of
{\it molecular bound states} of $m$ atoms.
 The appearance of the glass states ($T \le T_D$) is signaled by the fact that these
molecules still exist in the limit $ \lim_{ m \to 1} \lim_{w \to 0} \lim_{N \to \infty}$
(notice the order of limits). According to the above discussion, the ideal glass
 transition ($T_K$) is detected from the existence of a maximum of the  replicated
free energy $F_m=-\log(Z)/(\beta m)$ at a value of $m$ less than one. This is a well
defined mathematical problem, which fully describes our general strategy for 
computing the thermodynamics of the glass state. Of course this cannot be done
without resorting to some approximation schemes. We shall now develop one of them,
a kind of harmonic expansion in the solid phase, but several other
approximation schemes can be developed \cite{MPinprep}.

We
are interested in the regime of low temperatures where the molecules will have a
small radius, justifying a quadratic expansion of $v$  (we work here
with a regular potential $v(r)$, excluding hard cores).
We thus write the partition function in terms
of the center of mass and internal 
variables $z_i, u_i^a$, with $x_i^a=z_i+u_i^a$ and $\sum_a u_i^a=0$, expand the
energy to second order in $u$, and integrate over these quadratic fluctuations,
leading to:
\be
Z_m= {m^{Nd/2} \sqrt{2 \pi}^{N d (m-1)} \over N!} \int \prod_{i=1}^N dz_i
\exp\((-\beta m \sum_{i<j}
v(z_i-z_j) -{m-1 \over 2} Tr \log \((\beta M  \)) \))
\ee
where
the matrix $M$, of dimension $Nd \times Nd$, is given by:
\be
M_{(i \mu) (j \nu)}= \delta_{ij} \sum_k v_{\mu\nu}(z_i-z_k)-v_{\mu\nu}(z_i-z_j)
\ee
and $v_{\mu\nu}(r) =\partial^2 v /\partial r_\mu \partial r_\nu$ 
(the indices $\mu$ and $\nu$, running from $1$ to $d$, denote space directions).
We have thus found an effective Hamiltonian for the centers of masses $z_i$ of the 
molecules, which basically looks like the original problem at the effective
 temperature $T^*=1/(\beta m)$, complicated by the contribution of 
vibration modes. We shall proceed by using a 'quenched approximation', i.e.
 neglecting the feedback of
vibration modes onto the centers of masses. This approximation becomes
 exact close to the Kauzman temperature where $m \to 1$. The free energy is then:
\be
 {\beta F_m \over N} = -{d \over 2 m} \log(m)- { d (m-1) \over 2 m } \log(2 \pi)
-{1 \over m N} \log Z(T^*) +{m-1 \over 2 m} 
\la Tr \log \((\beta M  \)) \ra
\ee
where the partition function $Z(T^*)$ is simply that of the usual monatomic
liquid at the effective temperature $T^*$,
 and the expectation value
$\la . \ra$ is the Boltzmann expectation value at this same temperature.

Let us notice that the condition
for identifying the Kauzman temperature, $ {\partial \beta F_m \over \partial m} |_{m=1}=0$,
reads in our harmonic approximation:
\be
S_{liq}={d \over 2} \log(2 \pi e) - {1\over 2}\la Tr \log\((\beta M  \))\ra
\ee
$S_{liq}$ is the entropy of the liquid at the effective temperature
$T_{eff}$ which equals $T$ for $m=1$. The right hand side of 
this equation is nothing but the entropy $S_{sol}$ of an
harmonic solid with a matrix of second derivatives given by $M$. Thus we have 
found:
\be
{\partial \beta F_m \over \partial m }{\Biggr |}_{m=1}=S_{liq}-S_{sol}
\label{dfdm}
\ee
If $S_{liq}<S_{sol}$, one lies in the glass phase ($T<T_K$), while in the other
case $S_{liq}>S_{sol}$, the temperature is greater than $T_K$ (and of course less than $T_D$
if the spectrum of $M$ is positive). The complexity is then $S_c=S_{liq}-S_{sol}$,
as expected on general grounds \cite{remi}.

The harmonic expansion makes sense only if  $M$ has no negative eigenvalues,
which is natural since it is intimately related to the vibration modes of the glass.
 Notice that here we cannot describe activated processes,
and therefore we cannot see the tail of negative eigenvalues (with number decreasing as
$\exp(-1/T)$ at low temperatures), which is always present. It is known however
that the fraction of  negative eigenvalues of $M$ becomes negligible below
 the dynamical transition temperature $T_D$ \cite{INM1}.
So our harmonic expansion makes sense if the effective temperature $T^*$
is less than $T_D$.

Computing the spectrum of $M$ is an interesting problem of random matrix theory, in
a subtle case where the matrix elements are correlated. Some efforts have been devoted
to such computations in the liquid phase where the eigenmodes are called instantaneous
 normal
modes \cite{INM1}. It might be possible to extend these approaches to our case. Here
we shall rather propose a simple resummation scheme which should be reasonable
at high densities-low temperatures. Considering first the diagonal elements of $M$,
we notice that in this high density regime there are many neighbours to each point, and
thus a good approximation is to neglect the fluctuations of these diagonal
terms and substitute them by their average value. We thus write:
\be
\la Tr \log \((\beta M  \)) \ra =
N d \log (\beta   r_0 )+
\la Tr \log \(( \delta_{ij} \delta_{\mu\nu} - {1 \over r_0} v_{\mu\nu}(z_i-z_k) \)) \ra
\label{tracelog}
\ee
where
\be
r_0= {1 \over d} \int d^dr g(r) \Delta v(r)
\label{defr0}
\ee
and  $g(r)$ is the pair correlation in the liquid at the
effective temperature $T^{*}$.
In principle the spectrum at this stage still depends on {\it all} the correlation functions
of the liquid at $T^*$, as can be seen from an expansion of (\ref{tracelog})
in powers of $1/r_0$. A simple `chain' approximation involving only the pair correlation, consists
of approximating in each term of order larger than 2 in this expansion 
the full correlation by a product of pair correlations:
\bea
\int &dx_1&...dx_p \  g(x_1,....,x_p) \  v_{\mu_1 \mu_2}(x_1-x_2)...
v_{\mu_{p-1} \mu_p}(x_{p-1}-x_p )v_{\mu_p \mu_1}(x_p-x_1)
\\
&\simeq& 
\intk \((a(k)+{d-1 \over d} b(k)\))^p + (d-1) \intk  \((a(k)-{1 \over d} b(k)\))^p
\eea
where the functions $a$ and $b$ are defined by:
\be
\int d^d r \  g(r) v_{\mu\nu}(r) e^{ikr} \equiv \delta_{\mu\nu} \ a(k) +
\(( {k_\mu k_\nu \over k^2} -{1 \over d} \delta_{\mu \nu}\)) b(k)
\label{defab}
\ee
This chain approximation  selects  those contributions which survive in the high 
density limit, systematic corrections could probably be computed in the framework of the approach of 
\cite{Stratt}, we leave this for future work. Here and in what follows,
we have not written explicitly the
density: we choose to work with density unity and vary the temperature (density and temperature
variations are directly related in soft sphere systems onto which we focus below).

The free energy within the chain approximation is:
\bea
 {\beta F_m \over N} &=& -{d \over 2 m} \log(m)- { d (m-1) \over 2 m } \log(2 \pi)
-{1 \over m N} \log Z(T^*) + {d (m-1) \over 2 m } \log (\beta r_0)
\\
\nn
&+&
{(m-1) \over 2 m } \intk  \(( L_3 \(({a(k)+{d-1 \over d} b(k) \over r_0}\))
+
{(d-1)} L_3 \(({a(k)-{1 \over d} b(k) \over r_0}\)) \)) \\ 
&-&
{(m-1) \over 4 m } \int d^dr g(r) \sum_{\mu\nu} {v_{\mu\nu}(r)^2 \over r_0^2}
\label{chain}
\eea
where the function $L_3$ is defined as:
\be
L_3(x)=\log(1-x)+x+{x^2/2}
\ee
We can thus compute the replicated free energy $F_m$ soley from the knowledge
of the free energy and the pair correlation of the liquid
at the effective temperature $T^*$.
We have done this computation in the case of soft spheres
in three dimensions
with $v(r)=1/r^{12}$, using the free energy and pair correlation function of the liquid
given by the HNC approximation (obviously one could try to use better schemes of approximation
for the liquid, depending on the form of $v(r)$, in order to improve the
results; our point here is not to try to get the most precise results, but
 to show the feasibility of a quantitative  computation of glass properties
using the simplest approximations). We find (always at density unity)
a Kauzman temperature, obtained from
the vanishing of (\ref{dfdm}), which is $T_K \simeq .194$. When converted into the usual 
dimensionless parameter $\Gamma=\rho T^{-1/4}$, this gives $\Gamma_K \simeq 1.51$ which
is close to the glass temperature ($\Gamma \sim 1.6$)
 observed in simulations \cite{Han} with very fast cooling to avoid crystalization.
Simulations  done on 
binary mixtures (which do no crystallize) give a similar value for $\Gamma$. At the
level of precision we have now reached, one will need  both to 
do the theoretical computation for mixtures  and also to perform careful simulations in order
to disentangle the values of $T_K$ and $T_D$.

\begin{figure}
\centerline{\hbox{
\epsfig{figure=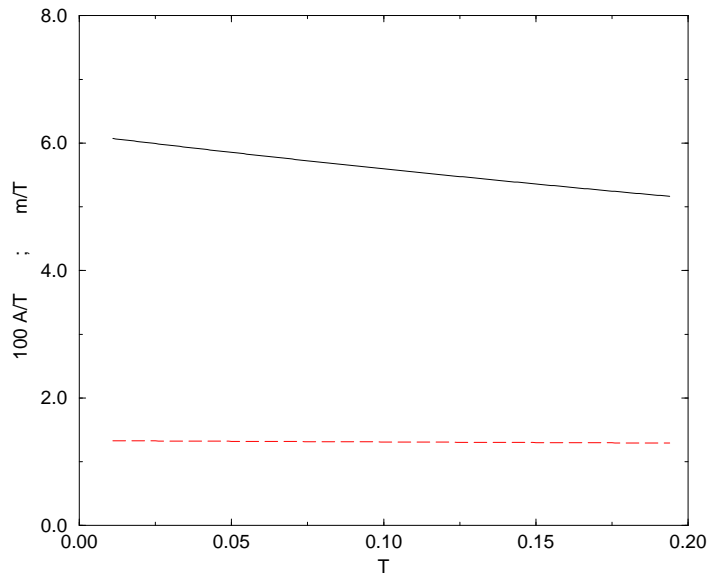,width=9cm,angle=-90}
}}
\caption{The full curve gives the inverse effective temperature $1/T^*= \beta m$ 
of the reference liquid as a function of
the temperature $T$ in the glass phase. The Kauzman temperature is  $T_K \simeq .194$. 
The dashed curves gives the square cage radius $A$ divided by the temperature and multiplied
by $100$. }
\label{fig1}
\end{figure}

 In fig. \ref{fig1} we show the values of the inverse effective temperature
($1/T^*= \beta*m$) and of the square cage radius $A$, defined as $A= {1 \over 3} (
\la x_i^2\ra -\la x_i \ra^2)$. This square cage radius has been obtained by using  in
 (\ref{Zm})
as attractive
potential: $w(r)=r^2/(4 \beta S)$, and differentiating the free energy:
\be
A= {2 \over d (m-1) N} \  {\partial (\beta F) \over \partial (1/S)} (S=\infty)
\ee

Notice that the effective temperature varies very little in the whole glass
phase and remains close to the Kauzman temperature, while the square cage radius is nearly 
linear in temperature in the whole glass phase, which is natural since non harmonic
 effects have been neglected. The value of $A$ at the Kauzman temperature is 
$A \sim 2.5 \ 10^{-3}$. This corresponds to a typical lateral displacement of the
particle in each direction of order $\sqrt A \sim .05$, which is $.045$ of the
mean interparticle distance, a value which gives the correct order of magnitude for the
 Lindeman ratio.
 
\begin{figure}
\centerline{\hbox{
\epsfig{figure=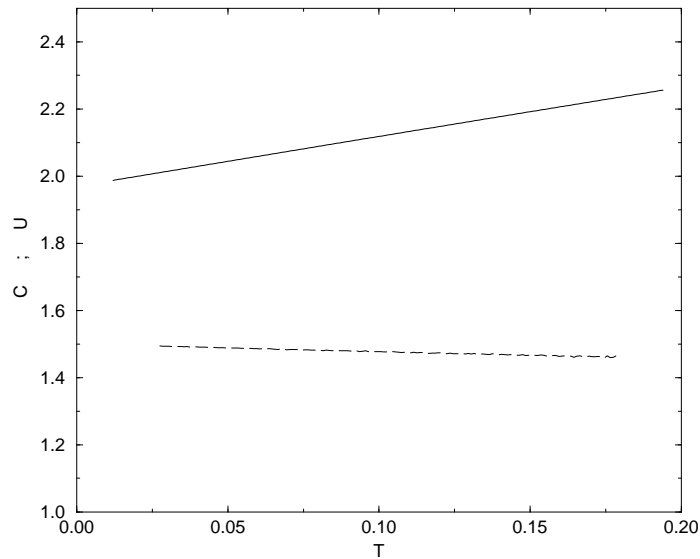,width=9cm,angle=-90}
}}
\caption{The internal energy (full line) and the  specific heat (dashed line)
 of the glass versus the temperature derived from our 
analytic computation. The specific heat value is close to the Dulong-Petit value of 3/2
and reaches this value in the zero temperature limit. }
\label{fig2}
\end{figure}

In fig. \ref{fig2} we give the result for the specific heat in the glass phase versus the
temperature. We see that it closely follows the Dulong-Petit law. This is the result that one
should obtain since we study a solid phase in the classical framework. Notice that it is not
at all trivial to derive this law from first principles in the glass phase. It is interesting
to see it coming out naturally from our computations: although we are basically using 
the properties of the liquid at the effective temperature $T^*$, the fact that the optimal number
of replicas $m$ vanishes linearly with $T$ at low temperatures naturally gives the Dulong-Petit law.

\begin{figure}
\centerline{\hbox{
\epsfig{figure=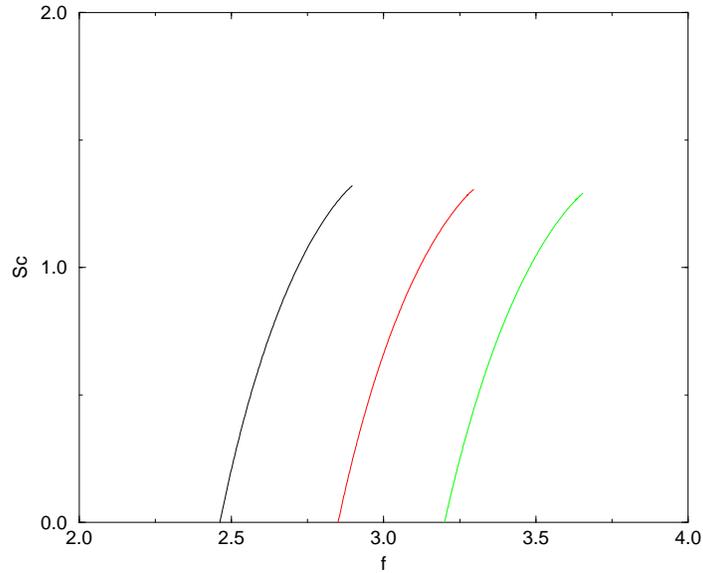,width=9cm,angle=-90}
}}
\caption{The configurational entropy $\Sigma(f)$ versus the free energy, 
computed analytically at temperatures
$T=.05,.1,.15$, from right to left. }
\label{fig3}
\end{figure}

  From the knowledge of $F_m$ as a function of $m$,
 we can compute the configurational entropy
as function of the free energy.
In fig. \ref{fig3}
we plot the result for $\Sigma(f)$ versus $f$ at three different temperatures. We see that the curves
are roughly parallel to each other, the main effect of the temperature changes being a shift in 
the $f$ axis.

\begin{figure}
\centerline{\hbox{
\epsfig{figure=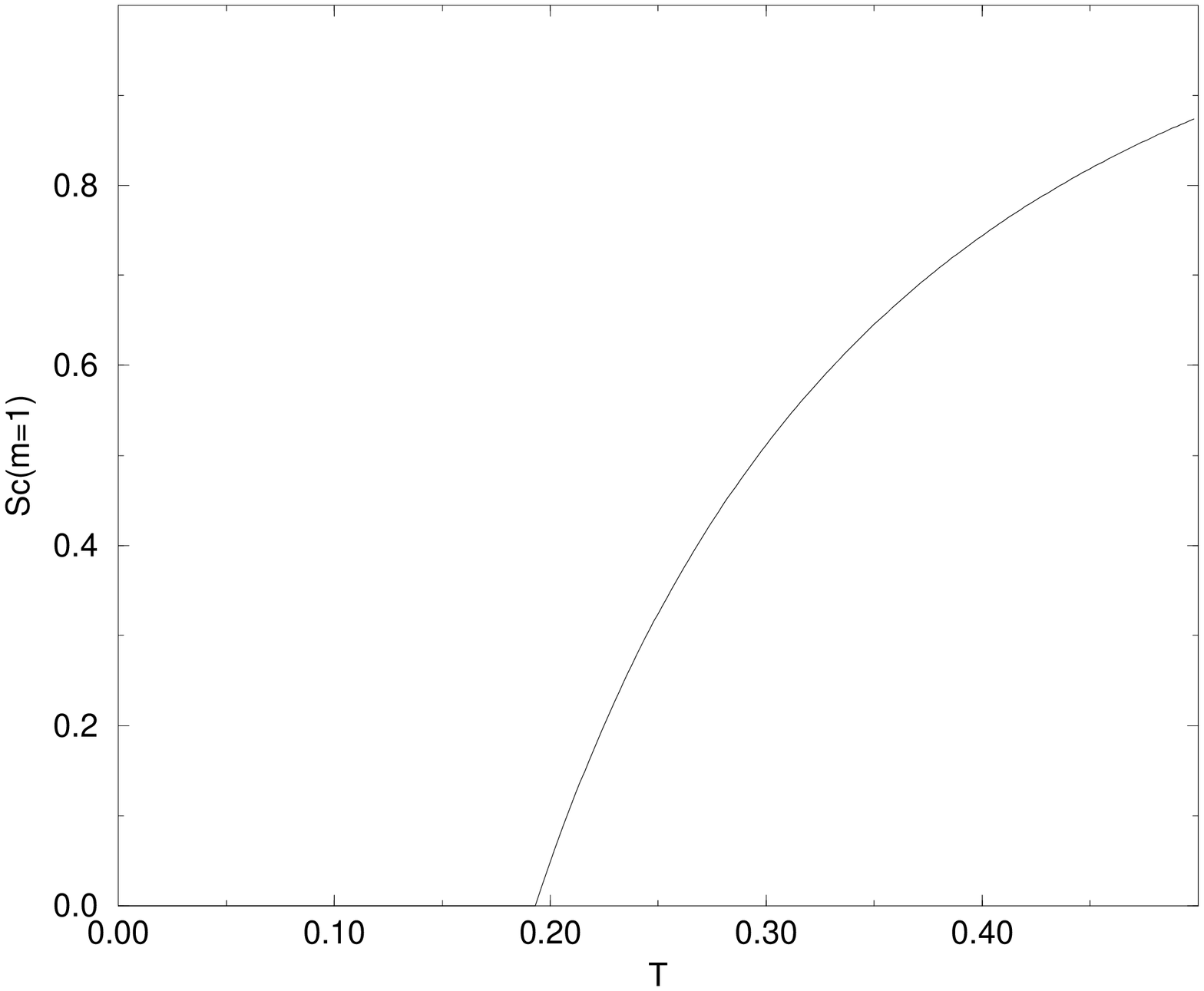,width=9cm,angle=-90}
}}
\caption{The equilibrium configurational entropy $\Sigma$ versus the 
temperature. It departs from zero above the Kauzman temperature. }
\label{fig4}
\end{figure}

As discussed above (see (\ref{SUM})), the value of the configurational
 entropy at equilibrium is
zero for $T<T_K$. It becomes non zero above $T_K$, where the saddle point in $m$ is at 
$m=1$. In fig. \ref{fig4} we plot the equilibrium configurational entropy versus
the temperature. It will be interesting to try to compare it
with numerical simulations.

To summarize, we have developed a well defined scheme for the analytic
study of the thermodynamics of the glass phase. The basic knowledge one
needs is the detailed  properties of the liquid (particularly
the instantaneous normal modes) close to the glass transition.
 We have shown that an
implementation of this scheme with rather simple approximations 
leads to very reasonable results. We hope to be able to refine these
approximations in a near future in order to get very precise predictions.
The extension of this approach to dynamical properties is also a
fascinating perspective.

\acknowledgements

We thank  A. Cavagna, D. Dean, I. Giardina,
and R. Monasson for very useful discussions.

\end{document}